\documentclass[namedreferences]{solarphysics}

\usepackage[hyperref,optionalrh,showbiblabels]{spr-sola-addons} 
\usepackage{graphicx}        
\usepackage{color}           
\usepackage{breakurl}        
\usepackage{multicol}

\newcommand{\aap}{    {\it Astron. Astrophys.}}

\newcommand{\apj}{    {\it Astrophys. J.}}

\newcommand{\grl}{    {\it Geophys. Res. Lett.}}

\newcommand{\jgr}{    {\it J. Geophys. Res.}}

\newcommand{\solphys}{{\it Solar Phys.}}
 
\newcommand{\ssr}{    {\it Space Sci. Rev.}} 

\newcommand{\araa}{ {\it Annu. Rev. Astron. Astrophys.}}
\newcommand{\planss}{ {\it Planet. Space Sci.}}
\chardef\us=`\_

\begin{document}
\begin{article}
\begin{opening}

\title{Near-Earth Interplanetary Coronal Mass Ejections and their association with DH type II Radio Bursts during Solar Cycles 23 and 24}
\author[addressref={aff1,aff2},corref,email={binalp@prl.res.in}]{\inits{B. D.}\fnm{Binal D.}~\lnm{Patel}\orcid{https://orcid.org/0000-0001-5582-1170}}
\author[addressref=aff1]{\inits{B. J.}\fnm{Bhuwan}~\lnm{Joshi}\orcid{https://orcid.org/0000-0001-5042-2170}}

\author[addressref={aff3,aff4}]{\inits{K.-S.}\fnm{Kyung-Suk}~\lnm{Cho}\orcid{https://orcid.org/0000-0003-2161-9606}}

\author[addressref={aff3}]{\inits{R.-K.}\fnm{Rok-Soon}~\lnm{Kim}\orcid{https://orcid.org/0000-0002-9012-399X}}

\author[addressref={aff5}]{\inits{Y.-J.}\fnm{Yong-Jae}~\lnm{Moon}
\orcid{https://orcid.org/0000-0001-6216-6944}}

\address[id=aff1]{Udaipur Solar Observatory, Physical Research Laboratory, Udaipur 313001, India}
\address[id=aff2]{Indian Institute of Technology, Gandhinagar 382355, India}

\address[id=aff3]{Space Science Division, Korea Astronomy and Space Science Institute, Daejeon 34055, Republic of Korea}
\address[id=aff4]{Department of Astronomy and Space Science, University of Science and Technology, Daejeon 34113, Republic of Korea}
\address[id=aff5]{School of Space Research, Kyung Hee University, Yongin 17104, Republic of Korea}
\runningauthor{B. D. Patel \textit{et al.}}
\runningtitle{ICMEs and their association with DH type II Solar Radio Bursts During Solar Cycles 23 and 24}

\begin{abstract}

We analyse the characteristics of interplanetary coronal mass ejections (ICMEs) during Solar Cycles 23 and 24. The present analysis is primarily based on the near-Earth ICME catalogue \citep{2010SoPh..264..189R}. An important aspect of this study is to understand the near-Earth and geoeffective aspects of ICMEs in terms of their association (type II ICMEs) versus absence (non-type II ICMEs) of decameter-hectometer (DH) type II radio bursts, detected by Wind/WAVES and STEREOS/WAVES. Notably, DH type II radio bursts driven by a CME indicate powerful MHD shocks leaving the inner corona and entering the interplanetary medium. We find a drastic reduction in the occurrence of ICMEs by 56\% in Solar Cycle 24 compared to the previous cycle (64 versus 147 events). Interestingly, despite a significant decrease in ICME/CME counts, both cycles contain almost the same fraction of type II ICMEs ($\approx$ 47\%). Our analysis reveals that, even at a large distance of 1 AU, type II CMEs maintain significantly higher speeds compared to non-type II events (523 km s$^{-1}$ versus 440 km s$^{-1}$). While there is an obvious trend of decrease in ICME transit times with increase in the CME initial speed, there also exists a noticeable wide range of transit times for a given CME speed. Contextually, Cycle 23 exhibits 10 events with shorter transit times ranging between 20-40 hours of predominantly type II categories while, interestingly, Cycle 24 almost completely lacks such ``fast" events. We find a significant reduction in the parameter $V_{\rm ICME} \times B_{\rm z}$, the dawn to dusk electric field, by 39\% during Solar Cycle 24 in comparison with the previous cycle. Further, $V_{\rm ICME} \times  B_{\rm z}$ shows a strong correlation with Dst index, which even surpasses the consideration of $B_{\rm z}$ and $V_{\rm ICME}$ alone. The above results imply the crucial role of $V_{\rm ICME} \times  B_{\rm z}$ toward effectively modulating the geoeffectiveness of ICMEs.


\end{abstract}

\end{opening}

\section{Introduction}

Coronal mass ejections (CMEs) consist of large-scale structures containing magnetised plasma expelled from the solar corona into the interplanetary medium. CMEs and their interplanetary counterparts (ICMEs) are known to be associated with a variety of other important phenomena, such as solar energetic particle (SEP) events, interplanetary (IP) shocks, geomagnetic storms (GS), which form essential ingredients of the current space weather research (\citealp[e.g.][]{2000JGR...10523153F, 2012LRSP....9....3W, 2017PhPl...24i0501C, 2017LRSP...14....5K}, \citealp{2021LRSP...18....4T}). The most energetic CME events may propagate the 1 AU distance within a day, while less energetic CMEs traverse the Sun-Earth distance in up to 4 days \citep{2005JGRA..110.9S15G, 2017SSRv..212.1159M}.

Observations of the solar wind confirm that CMEs at 1 AU generally have distinct plasma and field signatures by which they can be distinguished from the ordinary solar wind \citep{1990GMS....58..343G}. The characteristic features containing ICMEs in the solar wind flow include a stronger magnetic field, lower temperature of protons and electrons, lower pressure, and bidirectional electrons \citep{2003JGRA..108.1156C, 2006SSRv..123...31Z}. An interesting subset of ICMEs has enhanced magnetic field that rotates slowly through a large angle, so-called magnetic cloud (MC) \citep{1982JGR....87..613K}. Whenever the smooth rotating magnetic field signature is not observed, the ICME can be called ejecta \citep{2001JGR...10620957B}.


Halo CMEs propagating toward the Earth may produce geomagnetic storms when they collide with the Earth's magnetosphere \citep{2005JGRA..11011104K, 2007JGRA..11210102Z, 2017SSRv..212.1271K}. Geomagnetic storms produce a strong disturbance on the horizontal component of the global geomagnetic field, which is conventionally measured by the disturbance storm time index (Dst index). It is well established that the Dst minimum is directly linked to the southward component of the interplanetary magnetic field \citep[e.g.][]{1987P&SS...35.1101G, 2005GeoRL..3218103G, 2015JGRA..120.9221G}.

While travelling through the inner solar corona and interplanetary (IP) medium, energetic CMEs are known to generate shocks, which produce the slowly drifting feature in the radio dynamic spectra known as type II radio bursts \citep[e.g.][]{1963ARA&A...1..291W, 2001JGR...10629219G, 2007ApJ...663.1369R, 2018SoPh..293..107J}. The type II radio bursts observed at longer wavelengths (i.e., decameter-hectometer, kilometer regime) suggest the shock formation at higher heliocentric distances. The extension or origin of a type II radio burst in the decameter-hectometer (DH) domain implies the cases of stronger MHD shocks propagating from the inner corona and entering the IP medium \citep[e.g.][]{1998SoPh..183..165L, 2007ApJ...663.1369R, 2021SoPh..296..142P}. In this context, it is worth to note that the CMEs causing DH type II radio bursts have larger angular width and the majority are halo CMEs. Therefore, the study of ICMEs with respect to their association with DH type II radio bursts is extremely valuable to probe their ``invisible'' interplanetary propagation and possible geomagnetic consequences. 

In this article, we aim to broaden our understanding on the properties and propagation characteristics of ICMEs with respect to their association or lack with DH type II radio bursts during Solar Cycles 23 and 24. The present investigation becomes feasible due to an almost uninterrupted data set of IP type II radio bursts encompassing the two complete solar cycles. Further, the availability of a CME--ICME catalogue provides us with a unique opportunity for an in-depth exploration of interrelations between CMEs, ICMEs, and geomagnetic storms. In Section~\ref{sec:data_sources}, we present a brief description of the data sources for ICMEs, CMEs, and DH type II radio bursts. In Section~\ref{sec:analysis}, we provide a detailed analysis of various aspects related to CME--ICME associations. In Section~\ref{sec:conc}, we conclude with the results and findings of the present study. 

\section{Data Sources and Method}
\label{sec:data_sources}
For the present analysis, we have obtained data from the following catalogues. 
\begin{enumerate}
\item Near-Earth ICME catalogue\footnote{\url{www.srl.caltech.edu/ACE/ASC/DATA/level3/icmetable2.htm.}}:
The present analysis is based on the ICME catalogue, which contains a list of ICMEs detected by {\it in-situ} probes from May 1996 to December 2019 \citep{2003JGRA..108.1156C, 2010SoPh..264..189R}. The catalogue also comprises various ICME parameters such as disturbance time in solar wind profile, start and end time of ICMEs, the mean speed of ICMEs, temperature, the minimum value of geomagnetic Dst index during the entire time passage of ICME, and corresponding probable CME onset time based on LASCO observation. The criteria used for the identification of CME--ICME association is described in \cite{2003JGRA..108.1156C}. The catalogue also contains information about the observed ICME structure, i.e., MC or ejecta, based on the criteria proposed by \cite{1981JGR....86.6673B, 2001JGR...10620957B}. The catalogue also identifies complex events where the MC structure is not defined.  When making CME-ICME associations, the catalogue considered CMEs with angular extents of at least 100$^\circ$. \cite{2003JGRA..108.1156C} noted that for almost all the energetic events (i.e., bright and wide CMEs), one can make a direct link between the passage of the shock at the Earth and the time of a specific CME at the Sun. However, for a variety of reasons, the catalogue does not indicate a CME for many of the ICMEs. There can be several reasons for unability in identifying ICME--CME associations, for example, nonavailability of LASCO observations (i.e., a data gap) around the time when the associated CME might have occurred; LASCO observations show no evidence of a large CME during the several days prior to arrival of the ICME at Earth, likely due to the low density of the CME which falls below the LASCO detection threshold; the location of a CME is doubtful i.e., it originates near the limb or backside of the Sun and, therefore, unlikely to give rise to ICMEs at Earth. \cite{2003JGRA..108.1156C} provided a detailed discussions on difficulties in identification of CME--ICME associations. In the present work, we have considered only those events for which the CME-ICME association is clearly established in the catalogue.

\item Wind/WAVES type II bursts catalogue\footnote{\url{cdaw.gsfc.nasa.gov/CME\_list/radio/waves\_type2.html.}}: 
This catalogue contains the list of DH type II bursts identified by the Radio and Plasma Wave Experiment \citep[WAVES,][]{1995SSRv...71..231B} on board Wind spacecraft (1994--present) and SWAVE \citep{2008SSRv..136..487B} on board Solar TErrestrial RElations Observatory (STEREO; 2006--present). Wind/WAVES is the primary instrument providing continuous observations of low-frequency radio emission from the Sun in the frequency range of 14 MHz--20 kHz. SWAVE\footnote{\url{swaves.gsfc.nasa.gov/swaves_instr.html.}} has further complemented Wind/WAVES measurements by extending the frequency range up to 16~MHz. The combined observations from the two instruments are contained in the Wind/WAVES type II burst catalogue which is generated and maintained at the Coordinated Data Analysis Workshop (CDAW) data center by NASA and the Catholic University of America \citep{2019SunGe..14..111G}. The catalogue also identifies the CMEs associated with the DH type II radio bursts. 

\item Solar and Heliospheric Observatory (SOHO) Large Angle and Spectroscopic Coronagraph (LASCO) CME catalogue\footnote{\url{cdaw.gsfc.nasa.gov/CME\_list/.}}:
This catalogue contains the LASCO white light observations of solar corona from 2--30 R$_\odot$. This catalogue is generated and maintained at the Coordinated Data Analysis Workshop (CDAW) Data Center by NASA and the Catholic University of America in cooperation with the Naval Research Laboratory \citep{2009EM&P..104..295G} and provides important physical parameters that describe kinematic properties of CMEs, such as, linear speed, angular width, acceleration. We rely on the LASCO/CME catalogue to obtain the near-Sun (i.e., 2--30 R$_\odot$) parameters of CMEs associated with ICMEs arriving at 1 AU. 
\end{enumerate}

\begin{figure}
   \includegraphics[width=11 cm]{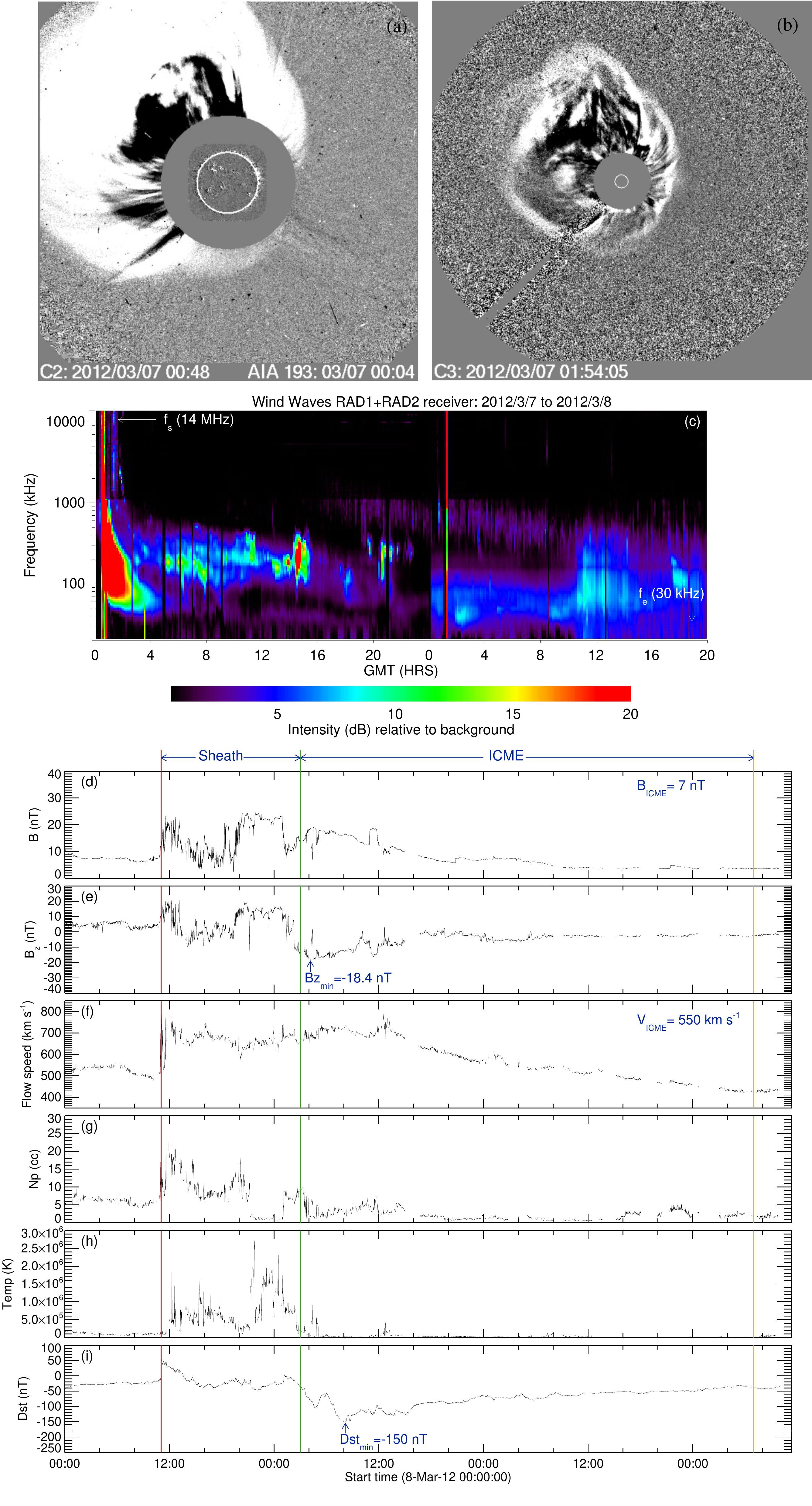}
   \caption{An example of a CME that exhibited DH type II radio burst, and ICME structure at 1 AU. Panels a and b represent LASCO C2 and C3 observations, respectively for the CME on 7 March, 2012. Panel c presents the dynamic spectrum of DH type II burst observed from the Wind/WAVES. Panels d--i show in-situ measurements of various parameters at 1 AU, \textit{viz.} $z$-component of magnetic field, flow speed, density, and proton temperature. Panel h represents the Dst value at near-Earth region. The vertical red, green, and orange lines represents arrival of shock at 1 AU, start, and end time of the ICME, respectively. The mean values for the magnetic field ($B_{\rm ICME}$) and speed ($V_{\rm ICME}$) of an ICME structure are annotated in the panels d and f. }
    \label{fig:example_cme_dh_icme}
\end{figure}

\begin{table}[h!]

\caption{Occurrence of ICMEs during Solar Cycles 23 and 24 for type II and non-type II ICME events.}
	\begin{tabular}{cccccc}

	\hline
		Solar Cycle & 	\multicolumn{3}{c} {Number of events} \\
      \cline{2-4}
            &Total ICME & Type II ICMEs & Non-type II ICMEs\\
    \hline
          SC 23 & 147& 68 (46\%) & 79 (54\%)\\
          SC 24 & 64 & 30 (47\%) & 34 (53\%) \\
          SC 23 + 24 &  211& 98 (46.5\%) & 113 (53.5\%)\\
          \hline
\end{tabular}
\label{table:occ_ICME_23_24}

\end{table}

\begin{table}[h!]
\caption{The results of two samples K-S test assessing the observed difference for various CME/ICME parameters between type and non-type II ICMEs for Solar Cycles 23 and 24. We consider the parameters for type II and non-type II ICMEs, separately. $N_1$ and $N_2$ represent the number of events for type II and non-type II ICMEs, respectively. The K-S test statistics and probability are denoted by D and Prob., respectively. For reference, we also provide the average values for the CME speeds ($V_{\rm CME}$), ICME speeds (V$_{\rm ICME}$), magnetic field (B), Dst index, southward magnetic field ($B_{\rm z}$), transit time, interplanetary (IP) acceleration, and -$V_{\rm ICME} \times  B_{\rm z}$.}

\begin{tabular}{cccccccc}
\hline
 & \multicolumn{7}{c}{Mean values} \\
\cline{3-4}
Solar Cycle & Parameters & Type II & Non-type II & $N_1$ &  $N_2$& D & Prob.\\
 \hline
23 &$V_{\rm CME}$ (km s$^{-1}$) & 1215 & 526 & 68 &79 &0.68 & 1.8$\times 10^{-15}$ \\
& $V_{\rm ICME}$ (km s$^{-1}$)  & 558 &  462 & 68 &79 & 0.35 & 1.3$\times 10^{-4}$ \\
& B (nT) & 12&10.8&68&79&0.135 &0.474\\
&$B_{\rm z}$ (nT) & $-$15.4&$-$12.7&68&79&0.17 &0.23\\
&Dst (nT)& $-$124.6 & $-$82& 68 &79 & 0.243 & 0.0212\\
&Transit time (hours) & 62.3 & 81.4 & 68 & 79 & 0.48 & 2.2$\times 10^{-8}$\\
&IP acceleration (m s$^{-2}$) & $-$3.72 & $-$0.29 & 68 & 79 & 0.683 & 4.81 $\times 10^{-16}$\\
&$-V_{\rm ICME} \times  B_{\rm z}$ (nT$\times$km s$^{-1}$) & 8917.5 &6078.1 &68 &79&0.25 &0.02\\
\cline{2-8}
24 &$V_{\rm CME}$ (km s$^{-1}$) & 1126 &  426 & 30 &34 &0.79 & 8.2$\times 10^{-10}$\\
& $V_{\rm ICME}$ (km s$^{-1}$)  & 488 &  418 & 30 & 34 & 0.39 & 0.0087\\

& B (nT) & 10.3&12.3&30&34&0.22 &0.38\\
&$B_{\rm z}$ (nT) & $-$11.9&$-$11.3&30&34&0.22 &0.35\\
&Dst (nT)& $-$80.2 & $-$61.7 & 30 &34 & 0.32 & 0.062\\
&Transit time (hours) & 71.1 & 90.3 & 30 & 34 & 0.51 & 2.7$\times 10^{-4}$\\
&IP acceleration (m s$^{-2}$) & $-$2.85 & $-$0.08 & 30 & 34 & 0.74 & 1.33 $\times 10^{-8}$\\
&$-V_{\rm ICME} \times  B_{\rm z}$ (nT$\times$km s$^{-1}$) & 5964.8 &4677.5 &30&34&0.27 &0.14\\
\hline

\end{tabular}
\label{table:ks_values}
\end{table}

Based on Near-Earth ICME catalogue, our primary dataset contains a total of 211 ICME events, which occurred in the period May 1996--December 2019. Theassociation between ICME and DH type II radio burst has been obtained by examining individual event listed in Near-Earth ICME and Wind/WAVES DH type II burst catalogues. In Table~\ref{table:occ_ICME_23_24}, we provide the statistics of the occurrence of ICMEs during Solar Cycles 23 and 24. We find that out of 211 ICMEs, 98 events (i.e., 47\%) produced DH type II radio bursts while the rest of the 113 events (i.e., 53\%) lacked DH type II radio bursts during their Sun-Earth propagation. Hereafter, we term the two subsets of ICMEs, defined on the basis of their DH type II association, as type II ICMEs and non-type II ICMEs.
\begin{table}[h!]
\caption{The results of two sample K-S test assessing the observed difference for various CME/ICME parameters between Solar Cycles 23 and 24. $N_1$ and $N_2$ represent the number of events for Solar Cycles 23 and 24, respectively. The K-S test statistics and probability are denoted by D and Prob., respectively. For reference, we also provide the average values for the CME speeds ($V_{\rm CME}$), ICME speeds (V$_{\rm ICME}$), Dst index, magnetic field (B), Dst index, southward magnetic field ($B_{\rm z}$), transit time, interplanetary (IP) acceleration, and $-V_{\rm ICME} \times  B_{\rm z}$.}

\begin{tabular}{ccccccc}
\hline
 \multicolumn{7}{c}{Mean values} \\
\cline{2-3}
  Parameters & Cycle 23 &Cycle 24 & $N_1$ &  $N_2$& D & Prob.\\
 \hline
$V_{\rm CME}$ (km s$^{-1}$) & 870 & 777 & 147 &64 &0.12 & 0.492 \\
 $V_{\rm ICME}$ (km s$^{-1}$)  & 510 &  453 & 147 &64 & 0.19 & 0.05 \\
 B (nT) & 11.4&11.3&147&64&0.08 &0.931\\
$B_{\rm z}$ (nT) & $-$13.9&$-$11.6&147&64&0.15 &0.22\\
Dst (nT)& $-$103 & $-$70.9& 147 &64 & 0.19 & 0.069 \\
Transit time (hours) & 72.7 & 81.3 &147& 64 & 0.17 & 0.11 \\
IP acceleration (m s$^{-2}$) & $-$1.74 & $-$1.38&147 & 64 & 0.09 & 0.76 \\
$-V_{\rm ICME} \times  B_{\rm z}$ (nT$\times$km s$^{-1}$) & 7247.3 &5202.1 &147&64&0.18 &0.07\\
\hline

\end{tabular}
\label{table:ks_values_cycle}
\end{table}

\begin{table}[h!]
\caption{Occurrence of Dst index associated with both ICMEs categories during Solar Cycles 23 and 24.}
\begin{tabular}{ccccccc}
\hline
Solar Cycle & Type&\multicolumn{4}{c}{Number of events}\\
\cline{3-6}
& &Total & Dst $\leq$ $-$100 nT & $-$100 nT $<$ Dst $\leq$ $-$50 nT &   $-$50 nT $<$ Dst \\
\hline
SC 23 & Non-type II & 79 &24 (30\%) & 24 (30\%) & 31 (40\%)\\  
&Type II & 68 &31 (46\%) & 24 (35\%) & 13 (19\%)\\
SC 24 & Non-type II &34& 6 (18\%) & 16 (47\%) & 12 (35\%)\\  
&Type II& 30&10 (33\%) & 11 (37\%) & 9 (30\%)\\

\hline
\end{tabular}
\label{table:dst_icme}
\end{table}

\begin{figure}
   \includegraphics[width=\textwidth]{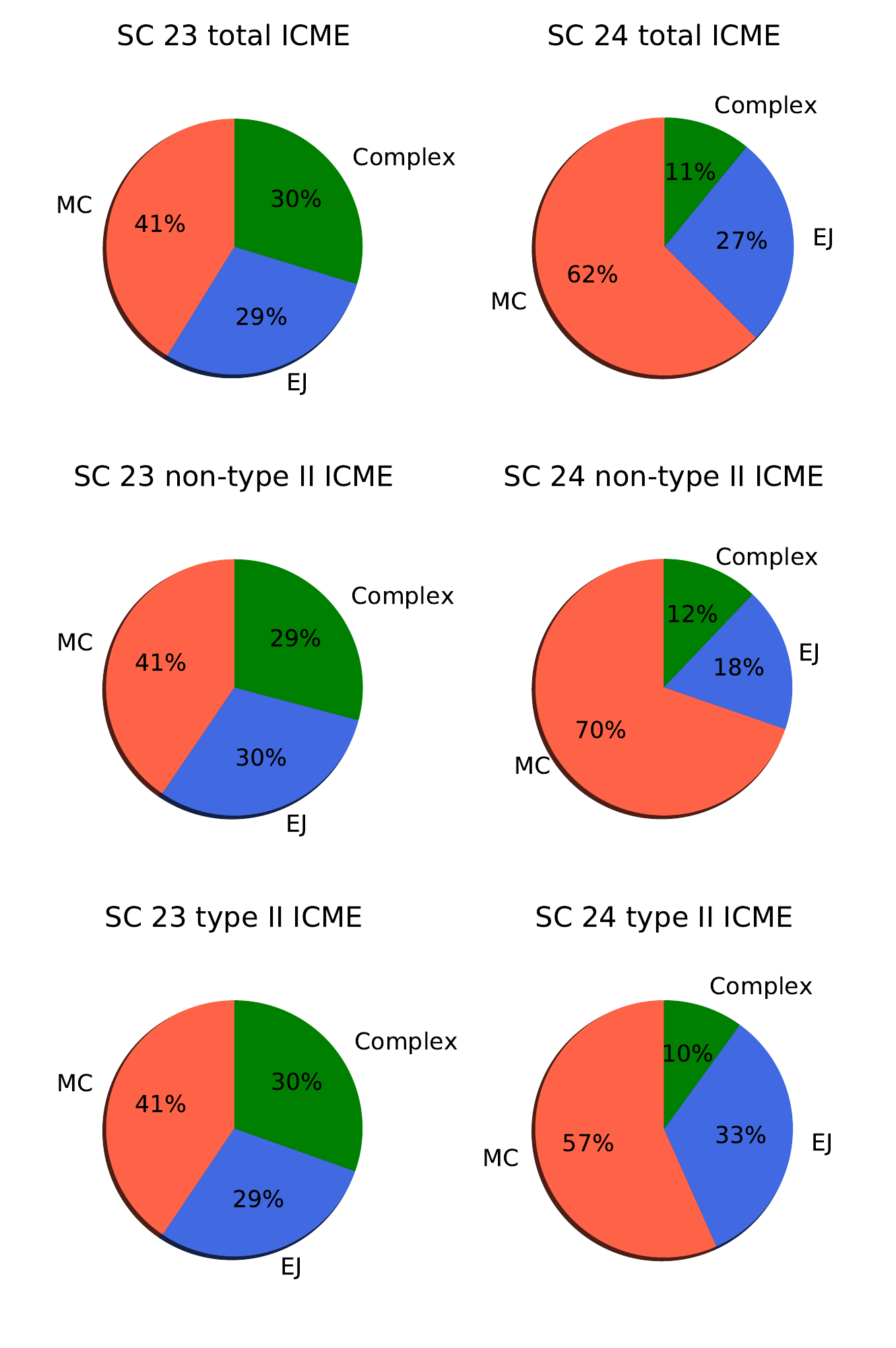}
   \caption{Pie chart showing the ICMEs associated with magnetic cloud (MC; red), ejecta (EJ; blue), and complex (green) events for total ICME events (top panel), non-type II ICMEs (middle panel), and type II ICMEs (bottom panel) during Solar Cycles 23 and 24. }
   \label{fig:pie_type_icme}
\end{figure}
In Figure~\ref{fig:example_cme_dh_icme}, we present an example of CME associated with the DH type II radio burst. Figures~\ref{fig:example_cme_dh_icme}a and b show a CME, observed on 7 March 2012 from LASCO/C2 and LASCO/C3 coronagraphs, respectively. In panel c, we plot Wind/WAVES radio dynamic spectrum during 7--8 March 2012, which shows DH type II radio burst corresponding to the CME of 7 March 2012 (see panels a and b) with starting and ending frequencies 14 MHz and 30 kHz, respectively. Panels d--i show the ICME parameters for the magnetic field, mean ICME speed, density, and temperature. The vertical black line shows the disturbance observed in the solar wind, which corresponds to the arrival of shock at 1 AU. The red and green vertical lines show the ICME start and end time, respectively. 

In order to compare the distributions of the observed CME and ICME parameters from the two samples (i.e., events under two specified categories), we have conducted a two sample Kolmogorov-Smirnov test (K--S test). The K--S statistic is defined as the largest absolute difference between the two cumulative distribution functions as a measure of disagreement \citep{1992nrfa.book.....P}. The key parameter of the test is the probability value between the cumulative distribution functions of two functions. A small value of probability ($\approx$ 0.05 or lower) suggests that the null hypothesis can be rejected, i.e., the two samples are drawn from two different distributions. Table~\ref{table:ks_values} presents the result of the K--S test applied to various CME/ICME parameters between type II and non-type II ICME categories for a given solar cycle. In Table~\ref{table:ks_values_cycle}, we assess the statistical significance of various CME/ICME parameters between Solar Cycles 23 and 24.
\begin{figure}
\includegraphics[width=\textwidth]{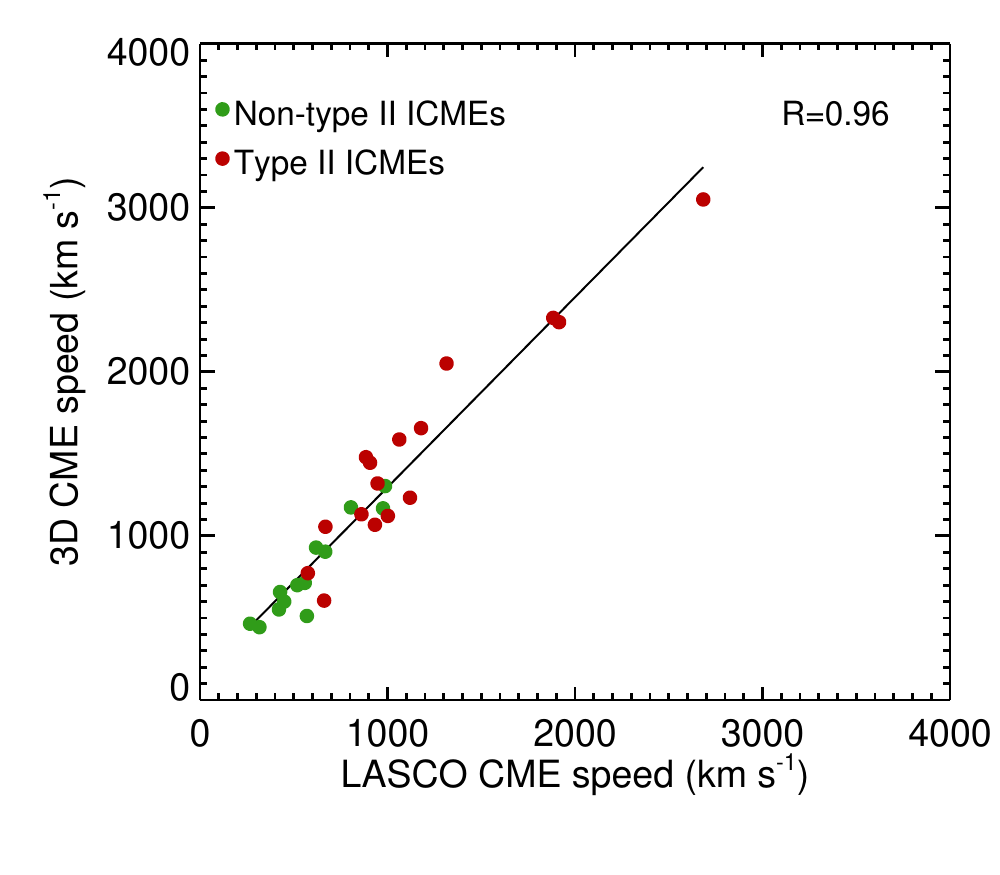}
\caption{Correlation plot between the LASCO CME speed (2D) and 3D CME speed for 29 events occurred during 2009-2013. These 29 events are common in our data-set and the catalogue by \cite{2016ApJ...821...95J}. The red and green symbols represent the type II and non-type II ICMEs, respectively. $R$ indicates the Pearson correlation coefficient for the linear regression line.}
\label{fig:speed}
\end{figure}
The early evolution of a CME, i.e., morphology, spatial expansion, and kinematics, depends upon several factors \citep[e.g.][]{2017SoPh..292..157B, 2017SoPh..292..118S, 2018ApJ...863...57B, 2020ApJ...900...23M}. The detailed case study by \cite{2017SoPh..292..157B} provides evidence that the kinks in the source active regions seem to be reflected in the expanded overlying flux rope structure that propagates into the heliosphere as a CME. \cite{2017SoPh..292..118S} explored the relative contributions of Lorentz force and aerodynamic drag on the propagation of CMEs for a set of 38 events. Their work revealed that the CMEs are subject to substantial acceleration or deceleration between 1$-$50 R$_\odot$. However, the consideration of these multiple factors in the CME propagation for individual events is beyond the scope of the present statistical investigations that explore broad characteristics of a large set of CME--ICME events for two complete solar cycles.  

To relate the near-Sun evolution of CMEs with their counterpart ICMEs at 1 AU, we rely on the CME speed, readily available from the CME catalogue based on LASCO observations (discussed earlier in this section). These CME speeds are essentially the linear speeds obtained by the height-time measurements from the LASCO CME white light coronographic images. We note that the LASCO CME speed provides us with a particular view of the CME projected on the plane of sky (POS). Hence, these LASCO 2D speeds contain inevitable projection effects, but the catalogue serves as the largest database which is definitely valid for any meaningful statistical analysis \citep[see review by][]{2021LRSP...18....4T}.

\section{Analysis and Results}
\label{sec:analysis}

\begin{figure}
\includegraphics[width=\textwidth]{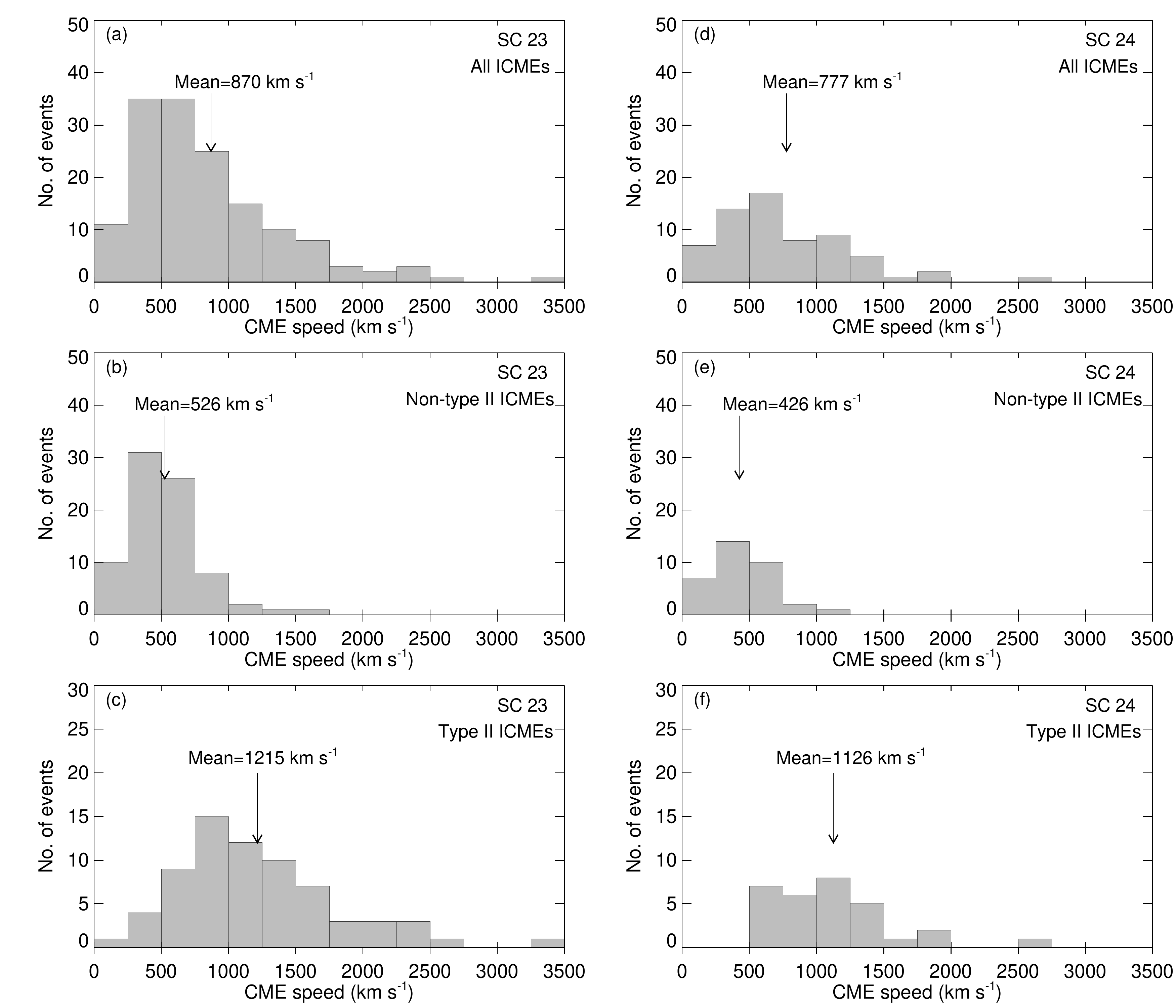}
\caption{Histograms showing the CME speed (within LASCO field of view) distribution for all ICME events (panels a and d), non-type II ICMEs (panels b and e), and type II ICMEs (panels c and f) during Solar Cycles 23 and 24. The mean CME speed value is annotated in each panel.}
\label{fig:histogram_speed_cme}
\end{figure}

\begin{figure}
\includegraphics[width=\textwidth]{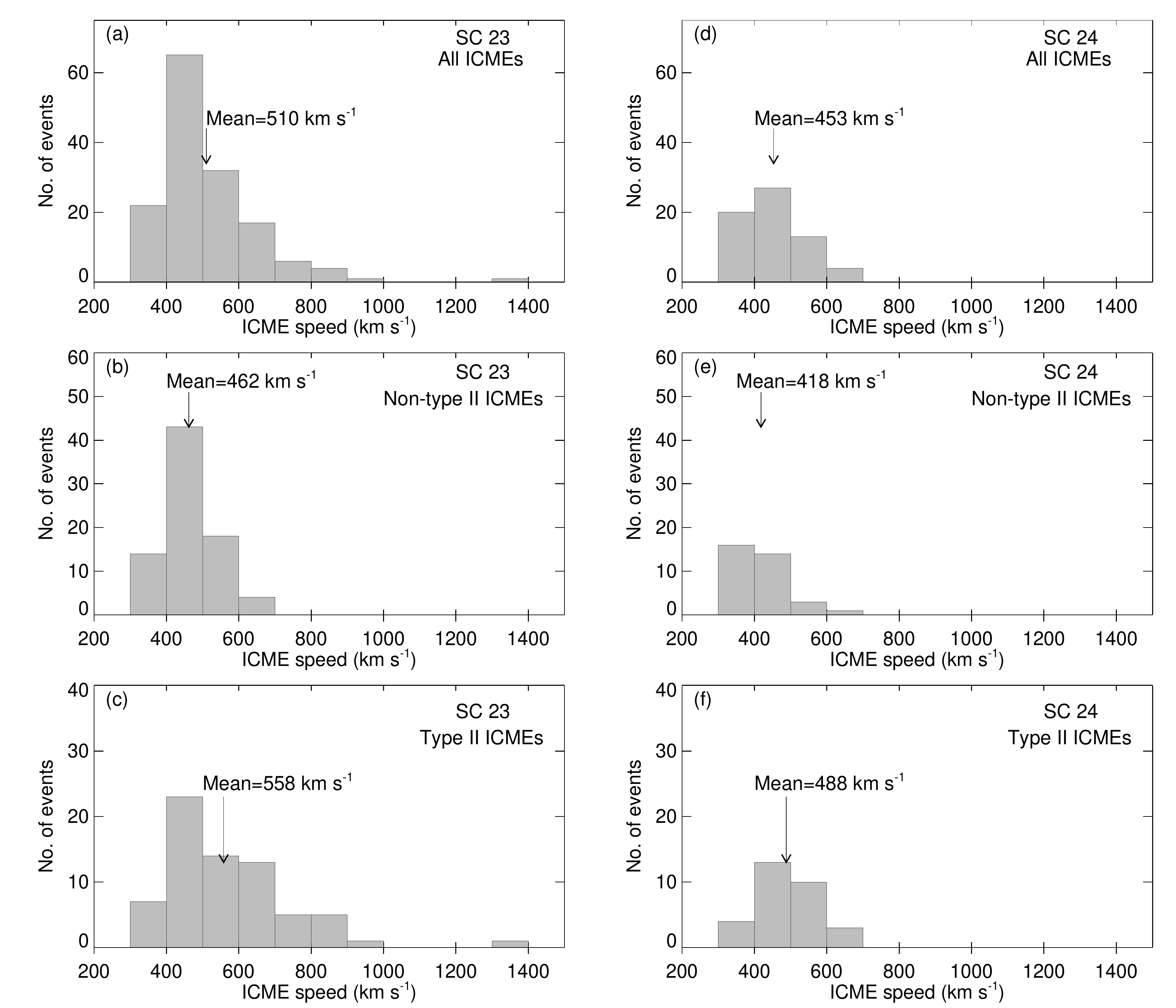}
\caption{Histograms showing the mean ICME speed distribution for all ICME events (panels a and d), non-type II ICMEs (panels b and e), and type II ICMEs (panels c and f) during Solar Cycles 23 and 24. The mean ICME speed value is annotated in each panel.}
\label{fig:histogram_speed_icme}
\end{figure}
\subsection{Overview of ICME Properties at 1 AU}
\label{sec:icme_overview}
In Table~\ref{table:occ_ICME_23_24}, we provide the statistics of the ICME events that occurred during Solar Cycles 23 and 24. Table~\ref{table:occ_ICME_23_24} further gives ICME counts in the two categories, i.e., type II and non-type II ICMEs. We note a drastic reduction (56\%) in the occurrence of total ICME events for Solar Cycle 24 with respect to the previous cycle (64 versus 147). It is interesting to note that almost the same proportion (47\%) of the total ICMEs produced DH type II radio bursts for both solar cycles (Table~\ref{table:occ_ICME_23_24}). Interestingly (and perhaps coincidentally) this fraction remains the same for both solar cycles. 

In Figure~\ref{fig:pie_type_icme}, we provide pie charts showing the occurrence of ICMEs as MC, EJ, and complex events, separately for Solar Cycles 23 (left panel) and 24 (right panel). The top, middle, and bottom panels show the total, non-type II, and type II ICMEs. Notably, for both ICME categories, the in-situ measurements detected the majority of magnetic cloud (MC) structures, suggesting the passage of full-fledged magnetic flux ropes embedded within the ICMEs \citep{1982JGR....87..613K, 2013SoPh..284...77K, 2015SoPh..290.1371M, 2019SoPh..294...54S}. Although Solar Cycle 24 is significantly weaker than the previous cycle in terms of overall ICME activity, it consists of a much higher fraction of MCs for both ICME categories. 

\begin{figure}
\includegraphics[width=\textwidth]{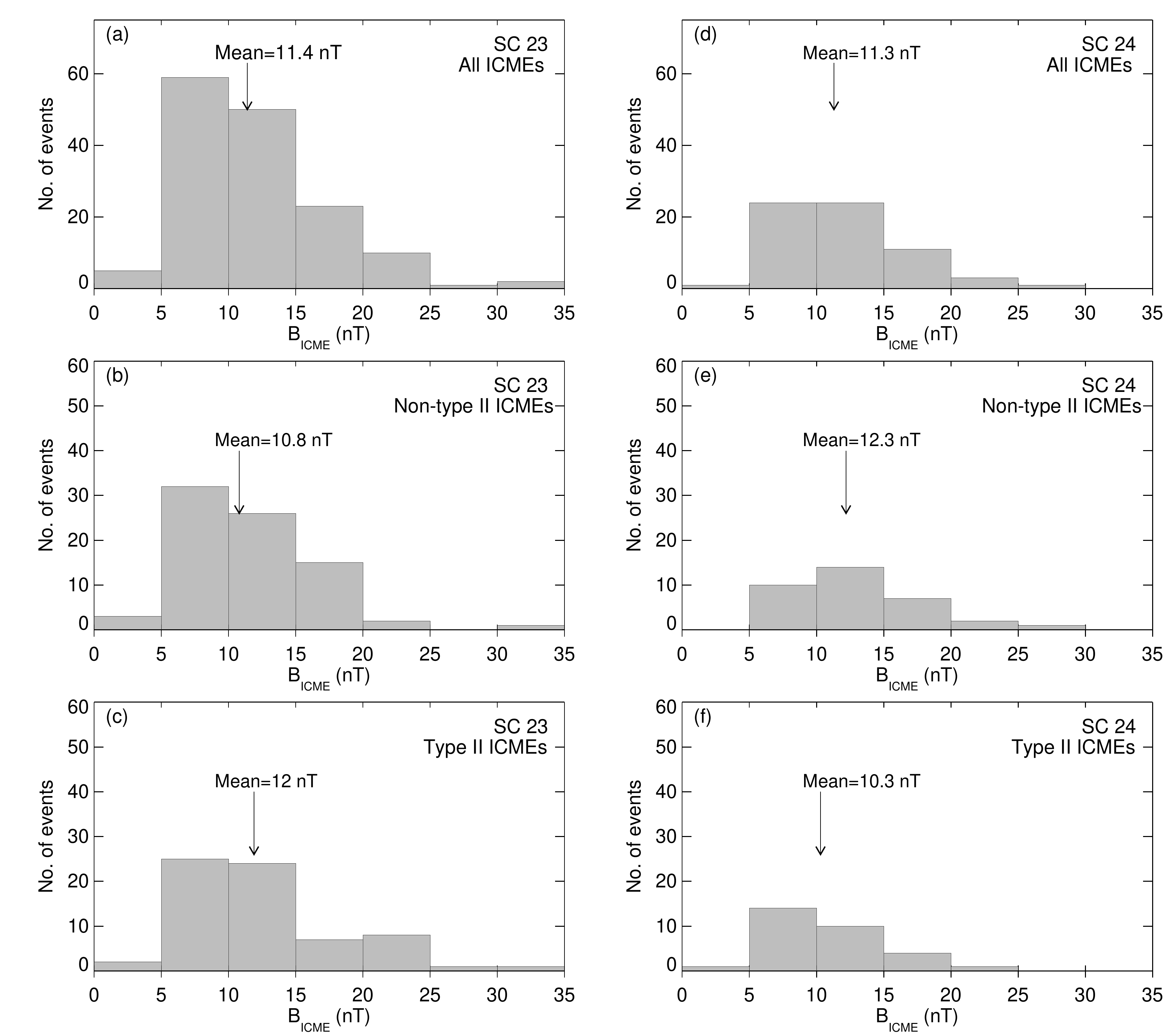}
\caption{Histograms showing the mean magnetic field ($B_{\rm ICME}$) distribution for all ICME events (panels a and d), non-type II ICMEs (panels b and e), and type II ICMEs (panels c and f) during Solar Cycles 23 and 24. The mean magnetic field value is annotated in each panel.}
\label{fig:histogram_mag_field}
\end{figure}

\begin{figure}
\includegraphics[width=\textwidth]{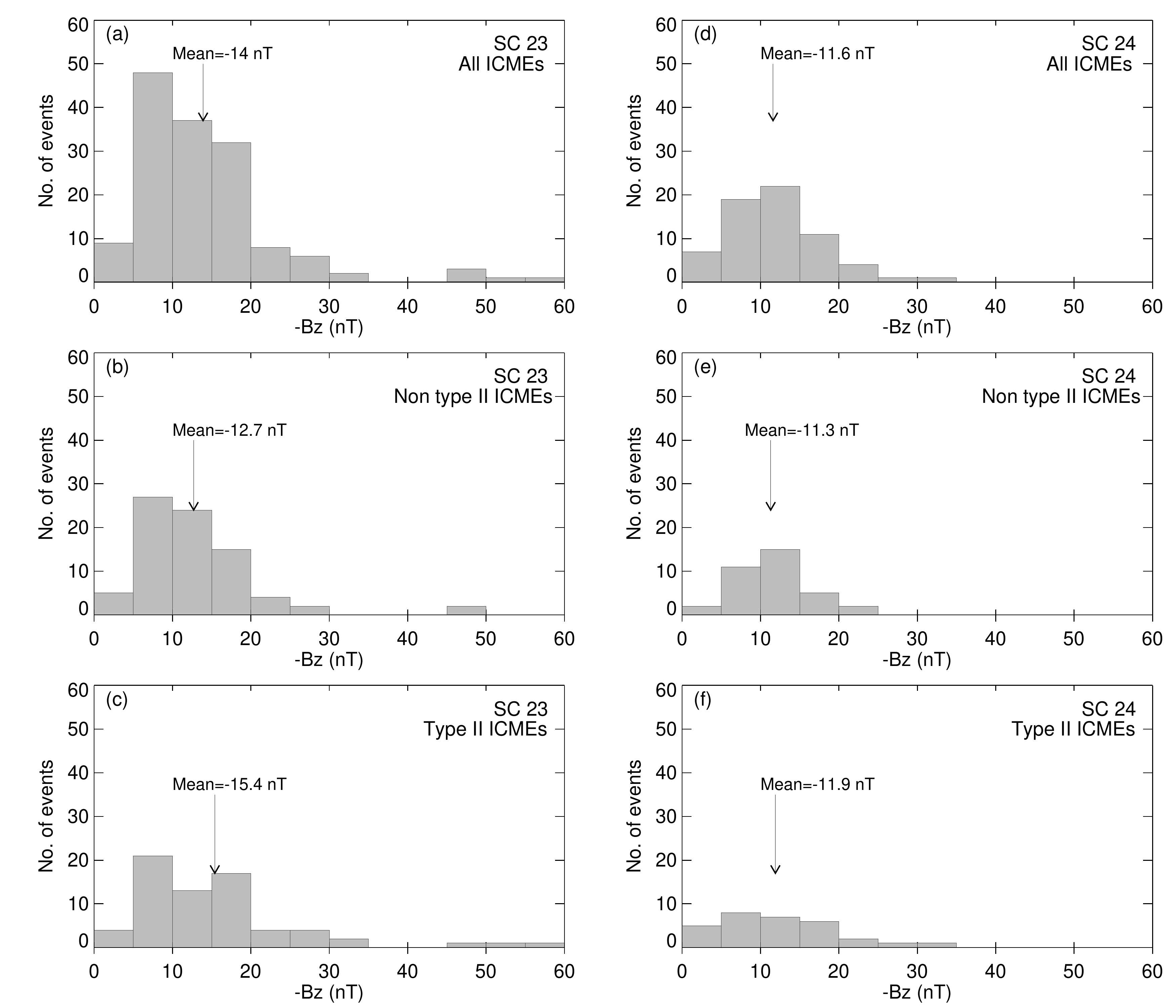}
\caption{Histograms showing the minimum southward magnetic field ($B_{\rm z}$) distribution for all ICME events (panels a and d), non-type II ICMEs (panels b and e), and type II ICMEs (panels c and f) during Solar Cycles 23 and 24. The mean value for $B_{\rm z}$ is annotated in each panel.}
\label{fig:histogram_bz}
\end{figure}

\begin{figure}
\includegraphics[width=\textwidth]{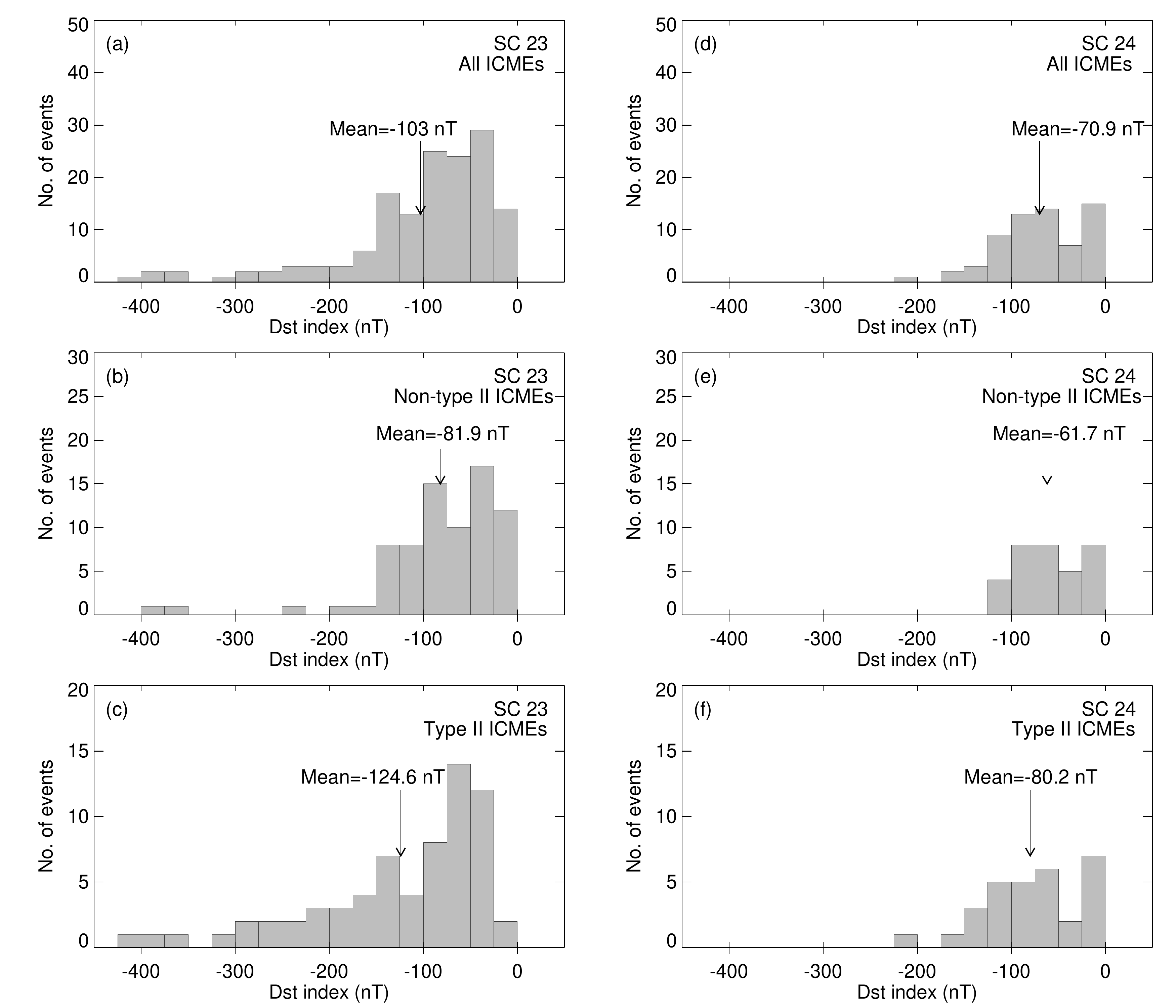}
\caption{Histograms showing the Dst index distribution for all ICME events (panels a and d), non-type II ICMEs (panels b and e), and type II ICMEs (panels c and f) during Solar Cycles 23 and 24. The mean value for the Dst index is annotated in each panel.}
\label{fig:histogram_dst}
\end{figure}
In Figure~\ref{fig:histogram_speed_cme}, we present the distributions of CME speeds associated with ICMEs during Solar Cycles 23 (left panel) and 24 (right panel). In order to assess the bias due to projection effects, we provide a comparison between the 3D CME speed and the LASCO speed in Figure~\ref{fig:speed}. Here, the 3D speeds are obtained from a catalogue\footnote{\url{ccmc.gsfc.nasa.gov/requests/fileGeneration.php}} of 3D CME parameters from 2009-2013 compiled by \cite{2016ApJ...821...95J}. Figure~\ref{fig:speed} presents the correlation between 2D and 3D speeds for 29 events, which are common in the present analysis and work of \cite{2016ApJ...821...95J}. The plot clearly reveals a good correlation between the two speeds, although projected speeds are underestimated in comparison to their 3D values. Earlier studies on CME kinematics using multiple-vantage point observations have revealed that the 2D speeds underestimate, on average, the 3D speeds by at least 20\% \citep{2016ApJ...821...95J, 2018ApJ...863...57B}. We therefore find that, although the projection effects are inevitable, the 2D CME speeds based on single viewpoint LASCO observations are apt for our statistical investigation. A comparison of histograms in Figure~\ref{fig:histogram_speed_cme} reveals that the mean CME speed is higher for events belonging to Solar Cycle 23 than those corresponding to the next cycle for both categories. The histograms further show a larger difference between the mean CME speeds for type II and non-type II ICMEs. The lower probability values of the K-S test statistics for both solar cycles suggest that the difference in CME speeds between type II and non-type II ICMEs is statistically significant (first row, Table~\ref{table:ks_values}), which essentially points toward the fact that type II associated CMEs possess higher kinetic energies \citep{2005JGRA..110.9S15G}. 

The histogram showing the distribution of ICME speeds measured at 1 AU is shown in Figure~\ref{fig:histogram_speed_icme} for Solar Cycles 23 (left panel) and 24 (right panel), separately. The histograms reveal that for both the categories, the mean ICME speed is higher for Solar Cycle 23 in comparison with Cycle 24. The histogram further reveals that for both solar cycles, the mean ICME speed for type II ICMEs is higher than that of non-type II ICMEs. We find that the difference in mean ICME speed for type II and non-type II ICMEs is higher for Solar Cycle 23 (96 km s$^{-1}$) than for Cycle 24 (70 km s$^{-1}$). The lower values of probability suggest that the difference of mean ICME speeds ($V_{\rm ICME}$) between type II and non-type II ICMEs is statistically significant for both solar cycles, with higher reliable difference for Solar Cycle 23 (second row, Table~\ref{table:ks_values}). We further find that for both solar cycles, the mean CME speed for type II ICMEs is almost twice in comparison with non-type II ICMEs. It is important to mention that the difference in CME speeds is much more pronounced at the near-Sun region than at 1 AU ({\it cf.} Figures~\ref{fig:histogram_speed_icme} and \ref{fig:histogram_speed_cme}). Interestingly, the application of the K-S test between the parameters of the two cycles reveals that the difference is statistically significant only for $V_{\rm ICME}$, while $V_{\rm CME}$ follows similar distributions (Table~\ref{table:ks_values_cycle}). 
\begin{figure}
   \includegraphics[width=\textwidth]{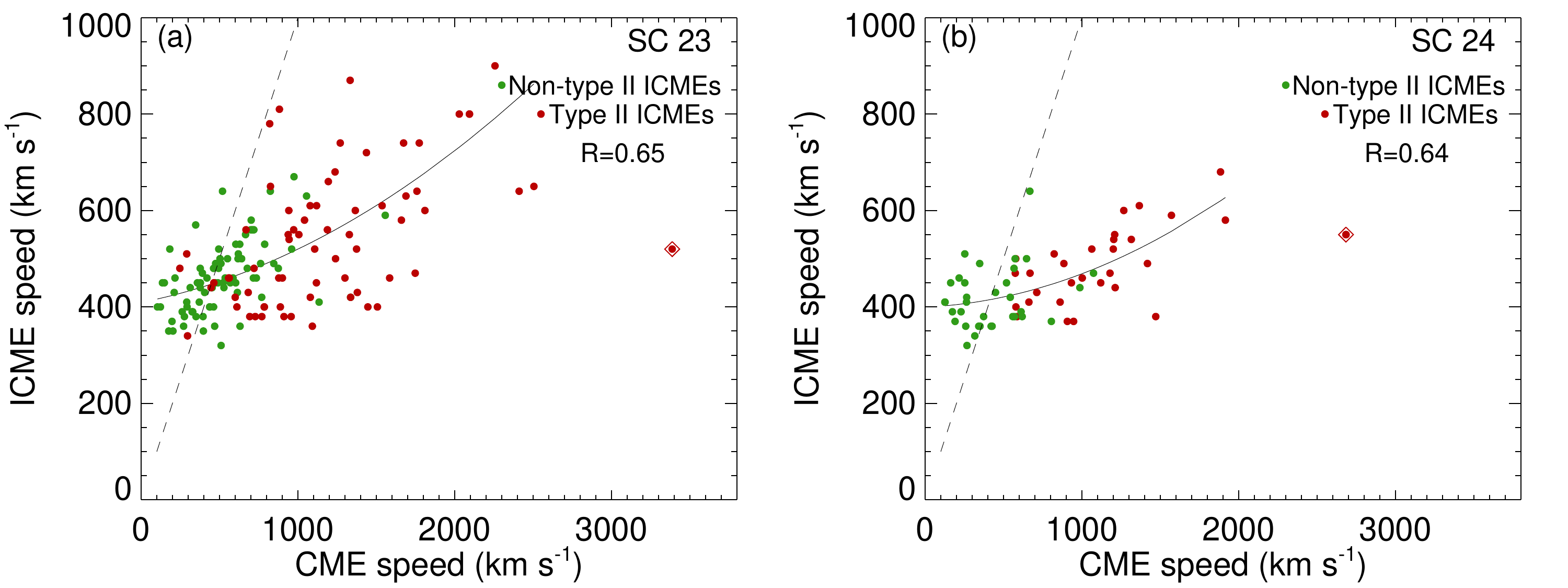}
   \caption{Relation between the initial CME speeds and the ICME speeds at 1 AU for Solar Cycles 23 (a) and 24 (b). The black curves indicate the quadratic least-squares fits to the data points. The dashed lines indicate the assumed constant velocity profiles for CMEs. The red and green symbols represent the type II and non-type II ICMEs, respectively. The data point shown with diamond symbol on each panel is an outlier and has been ignored during the quadratic least-squares fit. The quadratic correlation coefficient (R) for the quadratic fit is annotated in each panel.}
   \label{fig:speed_cme_icme}
\end{figure}
 
We analyse the distribution of the mean magnetic field ($B_{\rm ICME}$) for Solar Cycles 23 and 24, again for both categories of ICMEs in Figure~\ref{fig:histogram_mag_field}. The histograms reveal that the mean value of magnetic field does not show a statistically significant difference between the type II and non-type II ICMEs (third row, Table~\ref{table:ks_values}). This result likely indicates that, on average, all Earth arriving ICMEs possess a coherent, strong magnetic flux rope structure, irrespective of their near-Sun kinetic energies.

In Figure~\ref{fig:histogram_bz}, we provide the distribution for the minimum value of the southward component of magnetic field ($B_{\rm z}$) during ICME passage. The value of $B_{\rm z}$ is obtained from the OMNI database in GSM coordinate system at 1 minute time-resolution during the passage of ICME structure for each event. The histograms reveal that the mean value of $B_{\rm z}$ varies from $-$10 to $-$15 nT for different ICME categories over the two solar cycles. The histograms further reveal the higher mean absolute value of $B_{\rm z}$ for events of Solar Cycle 23 compared to the next cycle for both ICME categories. We find that for both solar cycles, the type II ICMEs have slightly higher values for the mean absolute value of $B_{\rm z}$ than the non-type II ICMEs. However, the higher probability values of the two samples K-S test for the minimum $B_{\rm z}$ between type II and non-type II ICME suggest statistically insignificant differences (fourth row, Table~\ref{table:ks_values}).

The histograms showing the Dst index associated with the ICMEs of different categories are presented in Figure~\ref{fig:histogram_dst}. We find that the mean value of Dst index is much lower for the events that occurred during Solar Cycle 23 ($-$103 nT) in comparison with Cycle 24 ($-$71 nT; Table~\ref{table:ks_values_cycle}). Our analysis shows a larger difference between the mean Dst index for type II and non-type II ICME categories for Solar Cycle 23, whereas the difference is smaller for Cycle 24. This result is confirmed by the lower value of the probability of the two samples K-S test for Dst index between type II and non-type II ICMEs which shows a statistically significant difference for Solar Cycle 23, whereas the difference is statistically insignificant for Solar Cycle 24 (fifth row, Table~\ref{table:ks_values}). Here, it is noteworthy that while the difference in mean $B_{\rm z}$ for type II and non-type II categories was found to be statistically insignificant (see Figure~\ref{fig:histogram_bz}), the difference in the mean values of Dst index for respective categories fall within the statistically acceptable range (Figure~\ref{fig:histogram_dst}; see also Table~\ref{table:ks_values} for K-S statistics); these observations imply complexities involved in the ICME-magnetosphere interaction that depend upon multiple ICME parameters \citep[][]{2010JGRA..11510111C, 2017SSRv..212.1271K}.

Based on the minimum value of Dst index, geomagnetic storms are classified as intense (Dst $\leq$ $-$100 nT), moderate ($-$100 nT $<$ Dst $\leq$ $-$50 nT) and weak \citep[Dst $>$ $-$50 nT;][]{1999SSRv...88..529G}. In Table~\ref{table:dst_icme}, we provide the statistics of the occurrence of geomagnetic storms associated with both ICME categories, separately for Solar Cycles 23 and 24. We find that type II ICME events show much higher association with intense geomagnetic storms. On the other hand, in terms of the occurrence of the weak and moderate geomagnetic storms, we cannot find any clear discrimination between type II and non-type II categories. Our analysis reveals that the type II ICMEs are prone to drive severe geomagnetic storms. We find that around 77\% (i.e., 76 out of 98) of type II ICMEs have produced moderate or intense geomagnetic storms. Further, only 62\% (i.e., 70 out of 113) events from non-type II ICMEs have produced intense or moderate geomagnetic storms.

\begin{figure}
   \includegraphics[width=\textwidth]{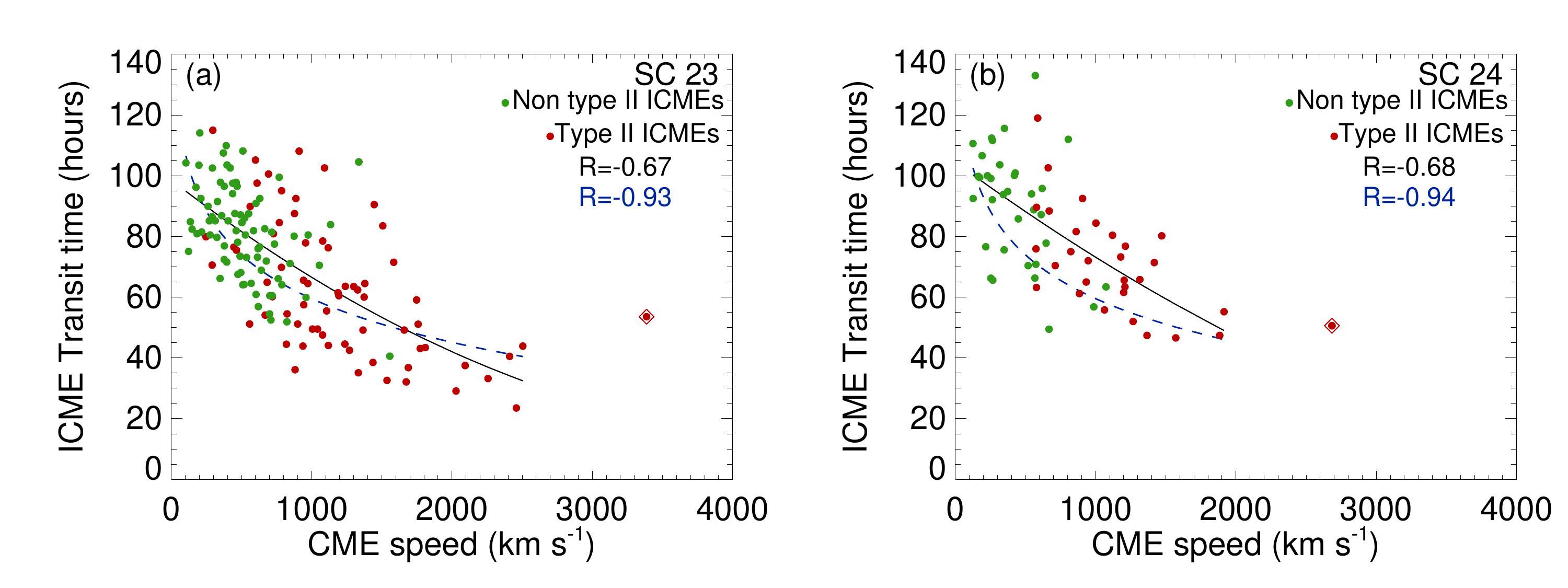}
   \caption{Relationship between the initial CME speeds ($V_{\rm CME}$) and the observed ICME transit-times ($T_{\rm ICME}$) for Solar Cycles 23 (a) and 24 (b). The red and green symbols represent the type II and non-type II ICMEs, respectively. The black curves indicate the quadratic least squares fits to the data points. The data point shown with diamond symbol on each panel is an outlier and has been ignored during the quadratic least-squares fit. The quadratic correlation coefficient (R) for the quadratic fit is annotated in each panel. The blue dashed curve indicates the fitting using a functional form obtained by \cite{2005AnGeo..23.1033S}, which is based on straight forward deceleration model assuming viscous drag.}
   
   \label{fig:vc_time}
\end{figure}

\subsection{CME-ICME Speed Relation}
In Figure~\ref{fig:speed_cme_icme}, we present the speed of ICMEs at the near-Earth region as a function of the initial CME speed for Solar Cycles 23 (panel a) and 24 (panel b). The data points corresponding to the type II and non-type II ICMEs are shown by red and green symbols, respectively. The dashed line in Figures~\ref{fig:speed_cme_icme}a and b indicates a constant velocity line (i.e., showing a hypothetical case of a CME with zero acceleration profile in the Sun-Earth journey). The plots suggest that for all ICMEs there exists a general trend in which the ICME speed increases with initial CME speed. We find that for both solar cycles, the ICME speed is higher for type II ICMEs in comparison with non-type II ICMEs. Further, we note that in Solar Cycle 24, the highest ICME speed is much lower in comparison to the previous cycle (700 km s$^{-1}$ versus 1300 km s$^{-1}$). It is important to mention that the ICME speeds exhibit a wide range for a given initial CME speed, suggesting that each CME has a different propagation profile which depends upon its energetics, interplanetary interactions among CMEs along with transient solar wind conditions. The best fit curves corresponding to the quadratic polynomials intersect the constant velocity line at 450 km~s$^{-1}$ and 420 km~s$^{-1}$ for Solar Cycles 23 and 24, respectively. This intersection value is nearly equal to the ambient solar wind speed and indicates the speed above which the deceleration of a CME is effective. We further note that the interaction values suggest low solar wind speed conditions during Solar Cycle 24 in comparison to Cycle 23.  
\begin{figure}
   \includegraphics[width=\textwidth]{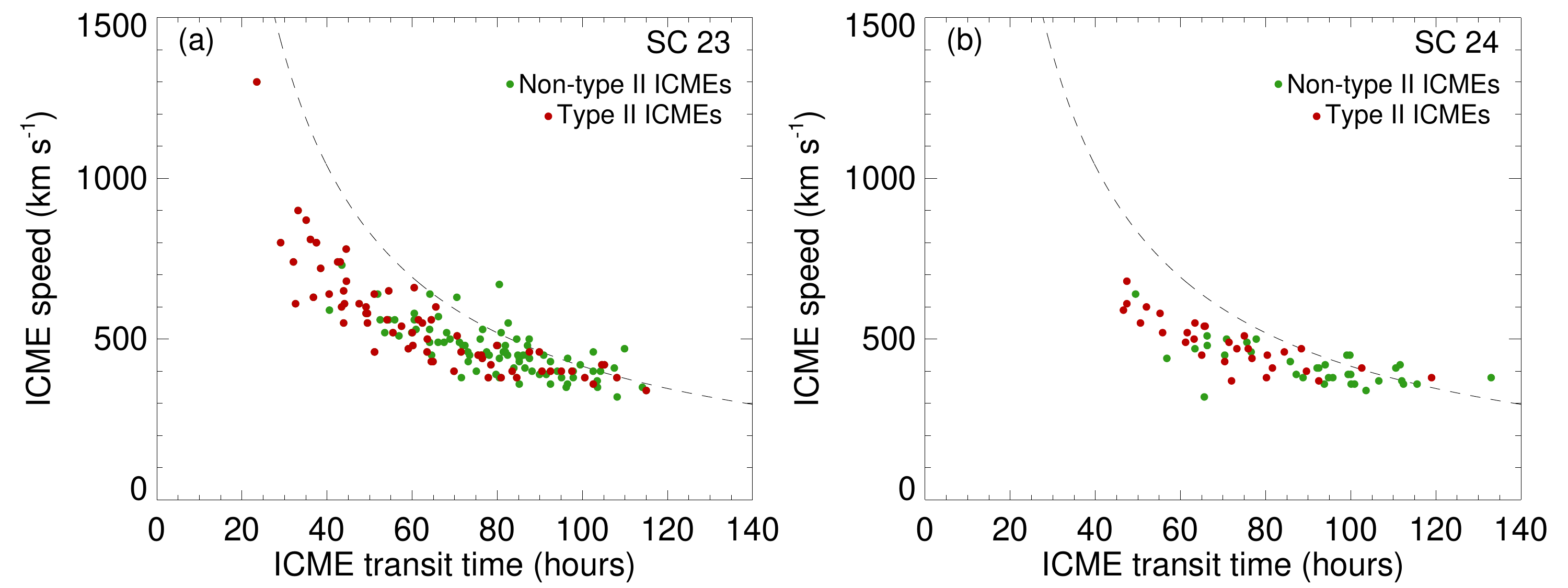}
   \caption{Relation between the ICME speeds and the observed ICME transit times for ICMEs for Solar Cycles 23 (a) and 24 (b). The red and green symbols represent the type II and non-type II ICMEs, respectively. The dashed curves show the estimated travel time assuming that CMEs propagated the Sun-Earth distance with a constant speed which is equal to their speed at 1 AU (i.e., ICME speed).}
   \label{fig:icme_transit_time}
\end{figure}
\subsection{Sun-Earth Transit Time of ICMEs}
The transit time of an ICME is estimated from the difference between the arrival of the ICME to the near-Earth region and onset time of the CME in LASCO C2 field of view. In Figure~\ref{fig:vc_time}, we plot the transit time for ICME events as a function of the initial speed of the CME for Solar Cycles 23 (panel a) and 24 (panel b). Data points corresponding to type II and non-type II ICMEs are denoted by red and green symbols, respectively. The solid lines in both panels represent the least squares quadratic fits obtained for all ICMEs. For Solar Cycle 23, the transit time ranges between 120 and 20 hours, while for Solar Cycle 24 the lower limit for transit time is at a significantly higher level, $\approx$ 45 hours. One event for each solar cycle is an outlier with the highest CME speed, and is not included in the quadratic fit. The outlier ICMEs are associated with CMEs dated on 10 November 2004 and 7 March 2012, for Solar Cycles 23 and 24, respectively. The least square fits to a second order polynomial to ICME transit time ($T_{\rm ICME}$) versus CME initial speeds ($V_{\rm CME}$) for Solar Cycles 23 and 24 are given by the following equations, respectively:
\begin{equation}
T_{\rm ICME}=98.6-0.04~V_{\rm CME}+3.7 \times 10^{-6}~V_{\rm CME}^2, 
\end{equation}
\begin{equation}
T_{\rm ICME}=110.6-0.04~V_{\rm CME}+4.9 \times 10^{-6}~V_{\rm CME}^2.
\end{equation}

The least square quadratic fits to the data suggest a decrease in the travel time as the initial CME speed increases. As mentioned in Section~\ref{sec:icme_overview}, the CMEs associated with type II ICMEs have higher speeds, and hence, they travel the Sun-Earth distance within a short duration. We find that the mean transit time for type II ICMEs is $\approx$ 62 and 71 hours for Solar Cycles 23 and 24, respectively. We further find that the mean transit time for non-type II ICMEs is $\approx$ 81 and 90 hours for Solar Cycles 23 and 24, respectively. In this context, it is worth mentioning that the transit time for both categories of ICMEs during Solar Cycle 24 is higher than the previous cycle. Further, we readily notice that the ICME transit times show a wide range for a given take-off speed of CMEs, which is consistent with earlier works that consider phases of the previous solar cycle besides some case studies (\citealp{2001JGR...10629207G}, \citealp{2004JGRA..109.6109M}, \citealp{2005AnGeo..23.1033S}, \citealp{2006SoPh..235..345M, 2007JGRA..112.5104K, 2010A&A...512A..43V}).

\begin{figure}
   \includegraphics[width=\textwidth]{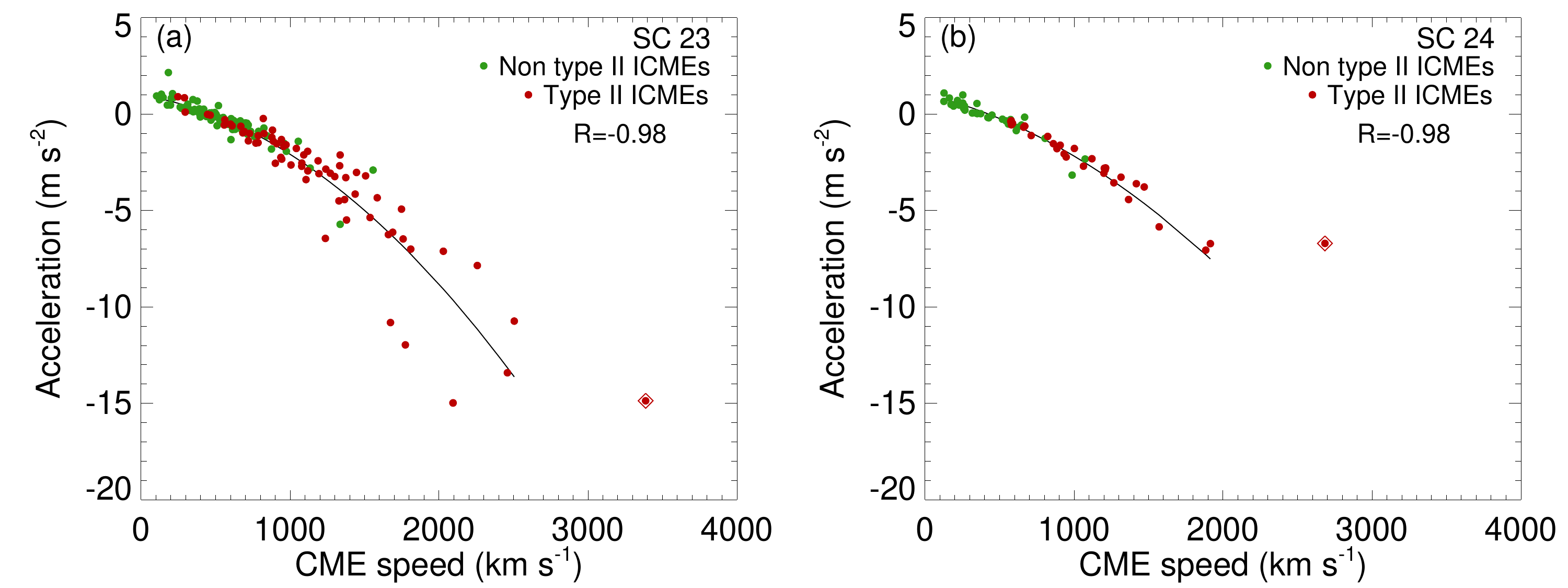}
   \caption{Relation between the CME speeds and the interplanetary accelerations for ICMEs for Solar Cycles 23 (a) and 24 (b). The red and green symbols represent the type II and non-type II ICMEs, respectively. The black curves indicate the quadratic least squares fit to the data points. The data point shown with diamonds are outliers and have been ignored during the quadratic least-squares fit. The quadratic correlation coefficient (R) for the quadratic fit is annotated in each figure.}
   \label{fig:speed_ip_acc}
\end{figure}
In previous studies, various empirical models have been constructed with different physical considerations on CME deceleration/acceleration to derive a plausible correlation between the CME radial speeds and their observed Sun--Earth travel times. \cite{2001JGR...10629207G} proposed an effective acceleration model to understand the statistical variations in CME transit times with respect to their initial speeds. Their model allows for the cessation of the interplanetary acceleration of a CME before 1 AU ($\approx$ 0.76 AU) while it propagates the remaining distance with a constant speed. Later \cite{2005AnGeo..23.1033S} considered a functional form on the basis of straight forward deceleration model assuming viscous drag \citep{2001SoPh..202..173V}. Their model leads to an asymptotic convergence of the CME speed to the ambient solar wind. In Figure~\ref{fig:vc_time}, the blue--dashed curve represents the optimum fit function derived by \cite{2005AnGeo..23.1033S} (see Equation 2 in their paper). We find that the function considering straight forward CME deceleration by viscous drag represents the relation between CME travel time versus initial speed very well. Based on the interplanetary scintillation (IPS) images of the inner heliosphere obtained from the Ooty Radio Telescope (ORT) for 30 ICMEs, \cite{2006SoPh..235..345M} found that up to a distance of 80 R$_\odot$, the internal energy of the CME (or the expansion of the CME) dominates and however, at larger distances, the CME’s interaction with the solar wind controls the propagation.
\begin{figure}
\includegraphics[width=\textwidth]{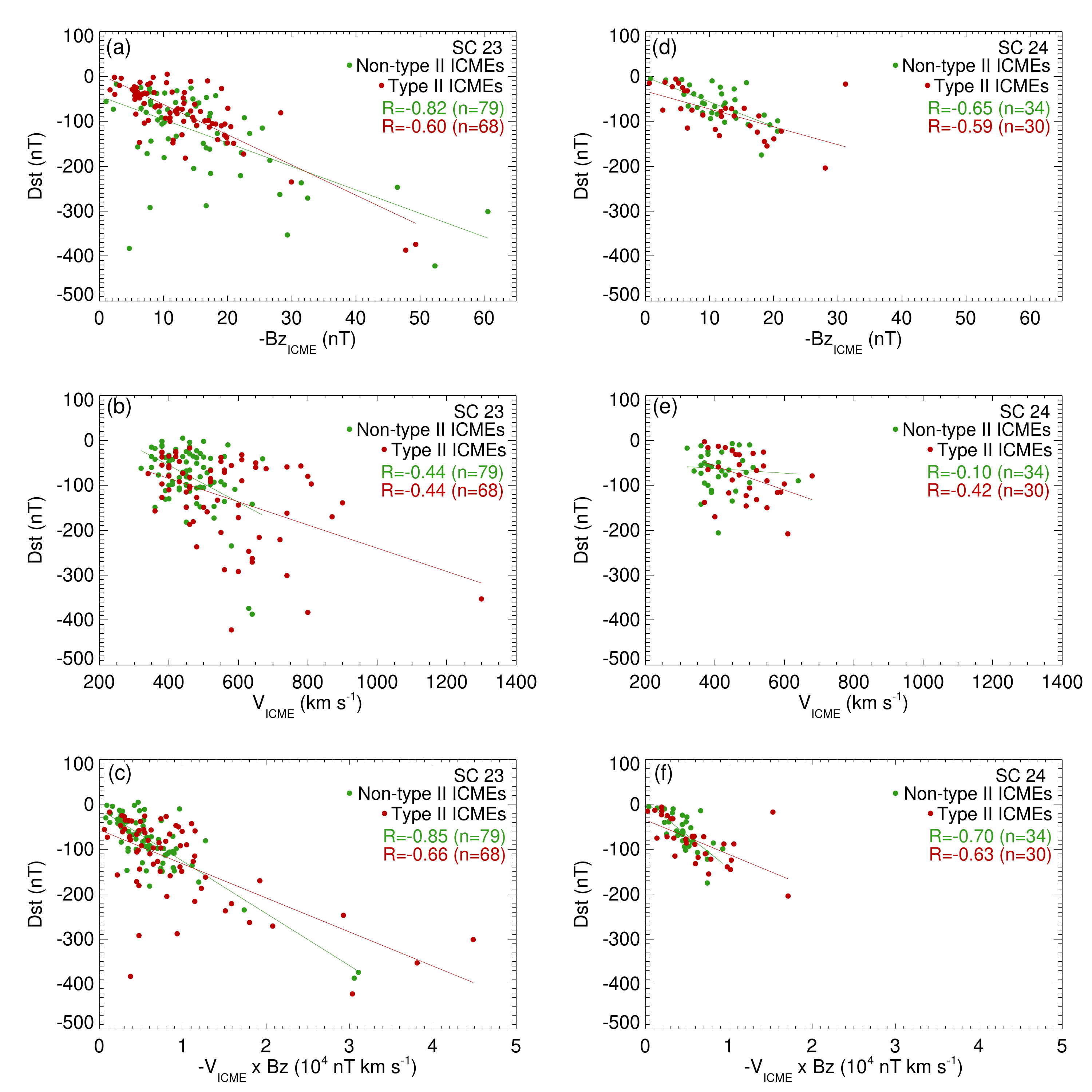}
\caption{Correlation plot between the minimum southward magnetic field ($B_{\rm z}$), ICME speed ($V_{\rm ICME}$), and -$V_{\rm ICME} \times  B_{\rm z}$ with the Dst index for Solar Cycles 23 (left panel) and 24 (right panel). The red and green symbols represent the type II and non-type II ICMEs, respectively. $n$ denotes the number of data points in each panel. $R$ in each plot indicates the Pearson correlation coefficient for linear regression line.}
\label{fig:corr_dst_merge}
\end{figure}
Comparing the CME transit time versus speed profiles for Solar Cycles 23 and 24 (see Figure~\ref{fig:vc_time}a and b), we find that Cycle 23 exhibits many events with shorter transit time ranging between 20--40 hours, while Cycle 24 completely lacks such events. For events of Cycle 24, the lack of shorter transit times primarily imply lower take-off speeds for CMEs. In this context, it is worth to note that CMEs during Cycle 24 were reported to show higher angular widths in comparison to the CMEs of Cycle 23 \citep{2014GeoRL..41.2673G} which would result into higher aerodynamic drag through the interplanetary medium \citep{2007A&A...472..937V}. Further, the transient changes in the interplanetary plasma and magnetic field conditions also play an important role in causing the difference in the interplanetary CME parameters (such as interplanetary expansion, transit time, and near-Earth ICME speed) during the different phases of solar activity \citep{2008GeoRL..3515105C}. Our analysis also reveals a significant difference in the Sun-Earth transit times for type II and non-type II events with shorter transit times for type II CMEs (Table~\ref{table:ks_values}). The fact that type II ICMEs have shorter transit time despite their wider structure at the near-Sun region implies that their internal energy effectively overpowers stronger aerodynamic drag. Thus, we find that the interplanetary evolution of a CME is highly influenced by the magnetic characteristics of the source active region where the magnetic flux rope structure develops and its early kinematic development in the near-Sun region \citep[e.g.][]{2008ApJ...677.1378K, 2010JGRA..11512108K, 2018JASTP.180...35G}.

In Figure~\ref{fig:icme_transit_time}, we explore the relation between the speed of ICMEs at 1 AU and their Sun--Earth transit time, separately for the events of Solar Cycles 23 (panel a) and 24 (panel b). In these plots, the dashed curve denotes the estimated travel time under the hypothetical assumption that all the CMEs propagate along the Sun-Earth distance with a constant speed which is equal to the ICME speed at 1 AU. The plots reveal that most of the ICMEs lie under the hypothetical transit-time curve which reflects the commonly accepted scenario in which the average speed of a CME during the Sun--Earth transit is higher than its speed at 1 AU. The existence of a few ICME events, predominantly of the non-type II category, lying on or above the dashed curve is noteworthy. Such events have probably undergone intermittent phases of acceleration during their interplanetary transit. 

In Figure~\ref{fig:speed_ip_acc}, we plot the effective acceleration of ICMEs through the interplanetary (IP) medium as a function of the initial speed of CMEs for Solar Cycles 23 (panel a) and 24 (panel b). Here IP acceleration for individual events is defined as the ratio between the difference in the initial (within the LASCO coronagraph) and final (at 1 AU) speeds of a CME, and corresponding Sun-Earth transit time. We note that the IP acceleration ranges from $-$15 to 2 m s$^{-2}$ for Solar Cycle 23, while the lower limit for deceleration is higher ($\approx$ $-$7 m s$^{-2}$) for Solar Cycle 24. The analysis further reveals that the type II ICMEs decelerate faster with CME initial speed for Solar Cycles 23 and 24. Notably, the CMEs launched with extremely high speed (i.e., the outlier events marked with a diamond symbol in Figure~\ref{fig:speed_ip_acc}) were subject to very high deceleration in the interplanetary medium. 

\subsection{ICMEs and Geomagnetic Storms}

The physical origins of geomagnetic activity are induction currents caused by the solar wind's electric field impacting the Earth's magnetosphere. They have two prime components: the ionospheric polar auroral electrojets and the near equatorial magnetospheric ring current, traditionally measured through Kp and Dst indices. We further note that Kp is not only made up of the intensity of the electrojet current systems, but also depends on their spatial position i.e., stronger compressions of the Earth's magnetosphere by the large pressure pulses driven by the solar wind cause movement of electrojets to lower latitudes, leading to strongly enhanced Kp values, but will appear less pronounced in the Dst index \citep[for detailed discussions see][]{2007swpe.book.....B}. In-situ measurements have revealed that ICMEs approaching the Earth (with speed $V_{\rm ICME}$) can possess long intervals of southward magnetic field ($B_{\rm z}$) which critically influence the energy transfer into the Earth's magnetosphere. The studies also point toward the fact that ICMEs cause relatively stronger response to the ring current index Dst, while sheaths produce a stronger response to high latitude auroral indices and to Kp (\citealp{2002JGRA..107.1121H, 2004AnGeo..22.1729H}; \citealp[see review by][]{2017SSRv..212.1271K}). Therefore, to relate the ICME arrival at 1 AU with its subsequent geomagnetic consequences, in this study we prefer to characterize storm strength by the Dst index.

To assess the geoeffectiveness of ICMEs, we primarily explore the relation of the Dst index, with $V_{\rm ICME}$ and $B_{\rm z}$. In Figures~\ref{fig:corr_dst_merge}a and d, we present the correlation between Dst index and $B_{\rm z}$ for non-type II and type II events for Solar Cycles 23 and 24, respectively. An examination of the correlation coefficients in different plots suggests a good correlation between the two parameters in all the cases. There is a noticeable difference in the correlation coefficients of Dst index versus $B_{\rm z}$ plots for non-type II and type II cases for Cycle 23 (0.85 versus 0.59) which is due to the wide spread in Dst values for a given $B_{\rm z}$, suggesting a much complex interplay of ICME and Earth's magnetosphere for individual type II associated events. The speed of the CME-flux rope at the time of interaction with Earth's atmosphere can be approximated by $V_{\rm ICME}$ measured by in-situ experiments at L1. The correlations of $V_{\rm ICME}$ with Dst index are presented in Figure~\ref{fig:corr_dst_merge}b and e, which suggest a weak (for Cycle 23) or no correlation (for Cycle 24). 

In Figure~\ref{fig:corr_dst_merge}c and f, we explore the correlation between Dst index and $V_{\rm ICME} \times  B_{\rm z}$. The quantity $V_{\rm ICME} \times  B_{\rm z}$ is essentially the dawn-to-dusk electric field imposed by the solar wind plasma and interplanetary magnetic field (IMF) on the Earth's magnetosphere (\citealp{1994JGR....99.5771G, 2015JGRA..120.9221G, 2018JGRA..123.6621R}; for a detailed discussion, see \citealp{2007swpe.book.....B}). Importantly, the parameter $V_{\rm ICME} \times  B_{\rm z}$ shows a strong correlation with Dst index for all the cases, which even surpasses the consideration of $B_{\rm z}$ or $V_{\rm ICME}$ alone. In this context, we note a significant difference of $V_{\rm ICME} \times  B_{\rm z}$ values between the type II and non-type II events for both cycles (Table~\ref{table:ks_values}). The synthesis of various correlation plots presented in Figure~\ref{fig:corr_dst_merge} indicates that a geoeffective ICME has to have an appropriate combination of its propagation speed and strength of the $z$--component of the magnetic field. It is important to note a significant reduction of $\approx$ 39\% in $V_{\rm ICME} \times  B_{\rm z}$ during Cycle 24 in comparison to the previous cycle. Thus, we find that the variation of solar wind electric fields has a more straight forward relation with the geomagnetic activity. 


\section{Conclusion}
\label{sec:conc}
In this article, we present a comprehensive statistical study of ICMEs that occurred during Solar Cycles 23 and 24. During the two cycles, a total of 211 Earth-reaching ICMEs were detected. The paper attempts to address different aspects of ICMEs: geoeffectiveness, variability at the near-Earth region, the Sun-Earth transit, and expulsion of CME from the coronal and near-Sun environment. In $\approx$ 47\% cases (98 out of 211 events), the fast expansion of CME produces signature in radio as DH type II radio bursts. As discussed in \cite{2021SoPh..296..142P}, the heliocentric distances up to which a CME associated coronal shock survive can be estimated from the end frequency of the corresponding type II radio burst \citep[see Figure 3 in][]{2021SoPh..296..142P}, and for about a quarter of events the shock even extends below hectometer frequencies ($\leq$ 200 kHz). Therefore, the DH type II associated CME-ICME events are of special interest in solar and solar--terrestrial studies.

A novel aspect of this study is to explore the interplanetary, near-Earth, and geomagnetic consequences of CMEs that produce DH type II emission against those that do not produce such radio bursts. Further, our study provides a comparison of the CME--ICME associations between Solar Cycles 23 and 24. We summarise the main findings of our study as follows:
\begin{itemize}
\item{With in-situ measurements, the majority of ICMEs are detected as clear magnetic cloud (MC) structures for type II and non-type II categories of ICMEs, which manifests towards the passage of full-fledged magnetic flux ropes through the observing spacecraft at 1 AU. Importantly, we observe a much higher fraction of MCs in Cycle 24 in comparison to the previous cycle (62\% versus 41\%).} 

\item{There is a noticeable difference in the mean CME speed for type II and non-type II CME categories (1215 km s$^{-1}$ versus 526 km s$^{-1}$ for Cycle 23; 1126 km s$^{-1}$ versus 426 km s$^{-1}$ for Cycle 24). Importantly, the type II CMEs remain much faster even at 1 AU over non-type II events (558 km s$^{-1}$ versus 462 km s$^{-1}$ for Cycle 23; 488 km s$^{-1}$ versus 418 km s$^{-1}$ for Cycle 24).}


\item{Although there is general and obvious trend that CMEs with the high take-off speed at the Sun tends to have shorter transit time, there is a wide range in transit times for a given initial CME speed. Notably, Cycle 23 exhibits several ICMEs (10 out of 147) with shorter transit time ranging between 20--40 hours, while Cycle 24 completely lacks such events.}


\item{The relation between Sun-Earth transit time of CMEs and their take-off speed is nicely represented by a straight forward deceleration model assuming viscous drag.} 

\item{In-situ measurements show comparable values of mean magnetic field and minimum southward component of the magnetic field associated with ICME (i.e., $B_{\rm ICME}$ and $B_{\rm z}$) for type II and non-type II categories, which implies that every Earth reaching ICME has to have a strong magnetic flux rope structure irrespective of their initial kinetic energy.}



\item{There is a significant reduction in $V_{\rm ICME} \times  B_{\rm z}$ during Solar Cycle 24 by 39\% compared to the previous cycle. Further, $V_{\rm ICME} \times  B_{\rm z}$ shows a strong correlation with Dst index, which even surpasses the consideration of $B_{\rm z}$ and $V_{\rm ICME}$ alone. Thus, we find that $V_{\rm ICME} \times  B_{\rm z}$ has more direct relation with the geomagnetic activity.}

\end{itemize}
In summary, our article investigates the Sun-Earth transit, near-Earth consequences and geoeffectiveness of ICMEs during Solar Cycles 23 and 24. Contextually, the two substructures of ICME that one typically observes at 1 AU $-$ sheath and magnetic cloud $-$ have different origin and solar wind conditions. In our subsequent work, we aim to study the separate contributions of the two ICME substructures as the driver of geomagnetic storms. Further, the identification of CME-forming magnetic flux ropes at the solar source region and the near-Earth environment from the unprecedented remote sensing and in-situ observations will provide important constraints to better understand the origin of geomagnetic storms.



\begin{acknowledgements}
We gratefully acknowledge the Near-Earth ICME catalog which forms the basis for the present study. We acknowledge the Wind/WAVES type II burst catalog and LASCO CME catalog. The LASCO CME catalog is generated and maintained at the CDAW Data Center by NASA and The Catholic University of America in cooperation with the Naval Research Laboratory. SOHO is a project of international cooperation between ESA and NASA. We acknowledge use of NASA/GSFC's Space Physics Data Facility's OMNIWeb service and OMNI data. We further acknowledge the SOHO, STEREO, ACE, and Wind missions for their open data policy. We are grateful to the anonymous reviewer of the paper for providing constructive comments and suggestions that have significantly enhanced the quality and presentation of the paper.	

\noindent
{\bf Disclosure of Potential Conflict of Interest} The authors declare that they have no conflict of interest.

\end{acknowledgements}

\bibliographystyle{spr-mp-sola}




\end{article}
\end{document}